\documentclass[aps,pra,twocolumn,showpacs]{revtex4}
\usepackage{bm}
\usepackage{amstext}
\usepackage{rotating}
\usepackage{graphicx}

\newcommand{\kbar} {\mathchar'26\mkern-9muk}

\begin{document}

\title{Nonlinear atom-optical delta-kicked harmonic oscillator using a
Bose-Einstein condensate}
\author{G. J. Duffy}
\author{A. S. Mellish}
\author{K. J. Challis}
\author{A. C. Wilson}
\affiliation{Department of Physics, University of Otago, P.O. Box 56, Dunedin, New
Zealand \\
}
\date{\today}

\begin{abstract}
We experimentally investigate the atom-optical delta-kicked harmonic
oscillator for the case of nonlinearity due to collisional interactions
present in a Bose-Einstein condensate. A Bose condensate of rubidium atoms
tightly confined in a static harmonic magnetic trap is exposed to a
one-dimensional optical standing-wave potential that is pulsed on
periodically. We focus on the quantum anti-resonance case for which the
classical periodic behavior is simple and well understood. We show that
after a small number of kicks the dynamics is dominated by dephasing of
matter wave interference due to the finite width of the condensate's initial
momentum distribution. In addition, we demonstrate that the nonlinear
mean-field interaction in a typical harmonically confined Bose condensate is
not sufficient to give rise to chaotic behavior.
\end{abstract}


\maketitle

The delta-kicked rotor is an extensively investigated system in the field of
classical chaos theory. During the last decade great progress has been
achieved in understanding quantum dynamics of a classically chaotic system
using atom-optical techniques and cold atoms. From an experimental point of
view, cold atoms in optical potentials \cite{Graham1992, Moore1995,
Anderson1998, Hensinger2000, Denschlag2002} provide an ideal environment to
explore quantum dynamics. To date, all experimental work has focused on
linear atomic systems, (see, for example, \cite{Hensinger2001, Steck2001,
Raizen1999, Duffy2004} and references therein) where the quantum dynamics is
stable due to the linearity of the Schr\"{o}dinger equation. In stark
contrast to the chaotic behavior of classical dynamics, the linear quantum
systems exhibit anti-resonance (periodic motion), dynamical localization
(quasi-periodic motion) or resonant dynamics \cite{Casati1979,
Shepelyansky1981}.

Recently, theoretical investigations have considered how the nonlinearity
due to many-body (collisional) interactions in a Bose-Einstein condensate
modifies the behavior of the atom-optical kicked rotor system, providing a
route to chaotic dynamics. Gardiner \textit{et al.} developed a theoretical
formalism to treat the one-dimensional nonlinear kicked harmonic oscillator
(a particular manifestation of the generic delta-kicked rotor) using
Gross-Pitaevskii and Liouville-type equations to describe the dynamics of a
Bose-Einstein condensate, and estimated the growth rate in the number of
non-condensate particles \cite{Gardiner2000}. Zhang \textit{et al.}
investigated the generalized quantum kicked rotor by considering a
periodically kicked Bose condensate confined in a ring potential for the
case of quantum anti-resonance \cite{Zhang2003}. As opposed to the familiar
periodic behavior exhibited by a corresponding linear system, they predicted
quasi-periodic variation in energy for a weak interaction strength and
chaotic behavior for strong interactions.

In this work we investigate the nonlinear delta-kicked harmonic oscillator
by performing experiments on Bose-Einstein condensates in a harmonic
potential. A Bose condensate of rubidium atoms tightly confined in a static
harmonic magnetic trap is exposed to a periodically pulsed one-dimensional
optical standing-wave potential. Our focus is on the particular case of
quantum anti-resonance for which the linear behavior is simple and well
understood \cite{Deng1999}. The finite width of the initial condensate
momentum distribution is shown to have a profound effect on the dynamics.
After a small number of kicks the behavior is dominated by dephasing of
matter wave interference. We present numerical solutions of the
Gross-Pitaevskii equation which match the observed behavior and confirm our
interpretation.

In the atom-optical kicked harmonic oscillator, the effective Planck's constant $\kbar$ can be adjusted to, in a sense, make the system ``more'' or ``less'' quantum
mechanical. At specific values of $\kbar$ - in
particular, where $\kbar$ is a rational multiple
of $2\pi$ - quite remarkable phenomena can occur in the form of so-called
quantum resonances and anti-resonances \cite{Izrailev1979, Izrailev1980,
Casati1984, Sokolov2000, Oskay2000, Daley2002,Duffy2004}. In this work we
focus our attention on the case of the $\kbar=2\pi$ anti-resonance at which the energy of a linear system exhibits simple periodic behavior. This anti-resonance
requires a particular initial momentum state which evolves in such a way
that during the period of free evolution in between kicks, the different
components of the state vector of the system experience a phase shift that
alternates in sign from one momentum component to the next, so that the
system returns identically to its initial state after every second kick. The
underlying physics of linear atom-optical kicking at anti-resonance has
already been neatly described, albeit in a different context \cite{Deng1999}. In the short pulse (thin grating) limit the first kick imprints a
sinusoidal phase profile onto the plane matter wave thereby populating a
number of momentum states (diffraction orders), and the phase evolution of
the $n^{th}$ state is proportional to $n^{2}$ so that after free evolution
(between kicks) corresponding to half the Talbot time ($T_{T}=h/4E_{r}$,
where the recoil energy $E_{r}=(\hbar k)^{2}/2m$, $k$ is the wave-vector and
$m$ is the atomic mass) the second pulse cancels the spatial variation
induced by the first. For multiple pulses this process repeats so that the
initial plane wave state is reconstructed after every second pulse.

Bose condensate evolution in an optical standing wave, or lattice, has
previously been well described by the Gross-Pitaevskii equation (GPE) (see,
for example, \cite{Blakie2000}), and condensate behavior in a kicked
harmonic potential can be described in this formalism using the one
dimensional GPE along the direction of the kicking beams,

\begin{eqnarray}
i \frac{\partial \psi (x,t)}{\partial t}&=&\left\{-\frac{\partial ^2}{\partial x^2}-\frac{\kappa}{\kbar \tau_p}\cos
\left(\sqrt{\frac{\kbar}{2T}}x\right)
\sum_{n=0}^{N}f(t-nT)\right.  \nonumber \\
&& \left.+\frac{1}{4}x^2+C|\psi(x,t)|^2\right\} \psi(x,t),
\end{eqnarray}

\noindent where $\psi (x,t)$ is the condensate wave function and
$\kappa =E_{r}\Omega T\tau _{p}/\hbar $ is the classical
stochasticity parameter (or kick strength) for the effective Rabi
frequency $\Omega $. Here $f(t-nT)$ represents a square pulse,
such that $f(t-nT)~=~1$ for $0~<~t-nT~<~\tau _{p}$, where $\tau
_{p}$ is the pulse length. The length scale is the characteristic
harmonic oscillator length $\sqrt{\hbar /2m\omega _{t}}$, and the
temporal scale is the effective trapping frequency $\omega _{t}$
along the axis of the kicking beams. The nonlinear strength
$C=(8\mu /3)^{3/2}$ is calculated such that the one dimensional
chemical potential $\mu $ is equal to the chemical potential of
the three dimensional condensate, in the Thomas-Fermi
approximation. Optimization techniques developed by Blakie and
Ballagh \cite{Blakie2000} are used to calculate the condensate
ground state and the GPE is evolved using a Runge-Kutta
fourth-order interaction picture algorithm
\cite{CaradocDavies2000}.

The energy of the system is calculated using

\begin{eqnarray}
E&=&\int \psi^* (x,t) \left\{-\frac{\partial ^2}{\partial x^2}\right.
\nonumber \\
&&-\frac{\kappa}{\kbar \tau_p}\cos\left(\sqrt{\frac{\kbar}{2T}}x\right)\sum_{n=0}^{N}f(t-nT)
\nonumber \\
&&\left.+\frac{1}{4}x^2+\frac{1}{2}C|\psi(x,t)|^2\right\} \psi(x,t) dx,
\label{energy}
\end{eqnarray}

\noindent which is evaluated after each kick to make a direct comparison
with experiment.

Note that in this formalism any non-condensate particles are not accounted
for. Previous theoretical papers \cite{Gardiner2000,Zhang2003} have
investigated the proliferation of non-condensate particles, and for our
nonlinearity, kicking strength, and number of kicks, this is predicted to be
negligible. Starting with a pure Bose condensate, we do not observe any
formation of non-condensate particles.

\begin{figure}[tbh]
\begin{center}
\includegraphics[width=\linewidth]{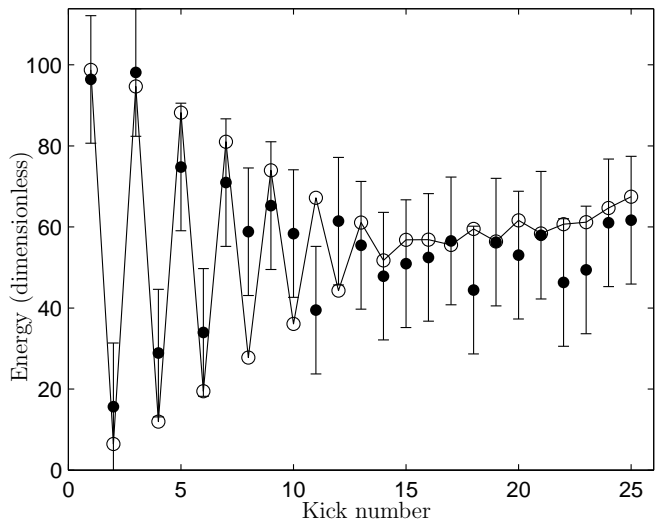}
\end{center}
\caption{Energy plotted versus kick number for the quantum anti-resonance
condition, $\kbar=2\pi$ with mean-field interactions. The solid circles are the measured mean energies
and the error bars include shot-to-shot variation and systematic uncertainty
in the calculation of the energy. The open circles are the corresponding
theoretical values computed using Eq.~\ref{energy} and the solid
line is to guide the eye.}
\label{fig:datatheoryline}
\end{figure}

Bose-Einstein condensates with up to ${10^{5}}$ $^{87}$Rb atoms
are created in the $F=2$, ${m_{F}}=2$ hyperfine state with no
discernible thermal component. A description of the BEC apparatus
was given previously \cite{Martin1999}, but there have been some
modifications. We now use injection-seeded diode lasers to drive
the two magneto-optical traps and atoms are transferred
continuously between the traps using a focused resonant laser
beam. Condensates are formed in the static harmonic potential of a
quadrupole-Ioffe-configuration trap \cite{Esslinger1998},
characterized by radial and axial oscillation frequencies of
$\omega _{r}/2\pi =164$~Hz and $\omega _{z}/2\pi =14$~Hz,
respectively. A condensate, while confined in the magnetic trap,
is then exposed to a pulsed optical standing wave generated by two
counterpropagating laser beams with parallel linear polarizations
derived from a single beam which is detuned 1.48~GHz from the
$5S_{1/2}$, $F=2$ $\rightarrow $ $5P_{3/2}$, $F^{\prime }=3$
transition. Each beam has an intensity of 1052~W/m$^{2}$ and
intercepts the condensate at an angle of 27$^{\circ }$ to the
radial direction. A double-pass acoustic-optic modulator is used
in each beam for switching the optical potential on and off. The
pulse length is 796~ns, which is much less than the minimum
classical oscillation period of 130~$\mu $s, so that the kicking
potential is well described as a thin phase grating
\cite{Ovchinnikov1999}. Following the kicks the momentum
distribution is determined from a time-of-flight absorption image
after a free expansion period of 29~ms by which time the momentum
components have separated. The energy of the atomic sample is
determined by calculating $(\int p^{2}dp)/2m$ then dividing by the
total number of atoms. The kicking period is 33.16~$\mu $s to
match the condition for the quantum anti-resonance at
$\kbar=8E_{r}T/\hbar \omega _{t}=2\pi $ (corresponding to half the
Talbot time), where $T$ is the pulse period in units of
$1/\omega_{t}$. The beam detuning and intensity were chosen to
give a relatively strong kicking strength while maintaining a
negligible spontaneous emission rate ($<$ 34~s$^{-1}$). Up to
twenty five kicks were delivered to the condensate for each
experimental run. For each number of kicks, the energy measurement
was repeated six times, and the mean value is plotted in
Fig.~\ref{fig:datatheoryline}.

 Figure~\ref{fig:datatheoryline} shows an experimental and
theoretical plot of energy versus kick number. The theoretical
calculation uses $\tau _{p}=6.9686\times 10^{-4}$, $T=0.029$ and
$C=50$ in correspondence with the experimental conditions. The
height of the optical potential has been adjusted ($\kappa=8.25$)
so that the energy after the first kick is consistent with the
experimental value. Initially, periodic behavior is observed, but
after several kicks the oscillation in the energy of the system
damps away to an average value (to within our experimental
uncertainty). The theoretical points indicate that this average
value gradually increases, but no further significant oscillation
is expected. This steady increase occurs because in the time
between kicks, atoms moving in the harmonic potential gain a small
amount of potential energy.

\begin{figure}[tbh]
\begin{center}
\includegraphics[width=\linewidth]{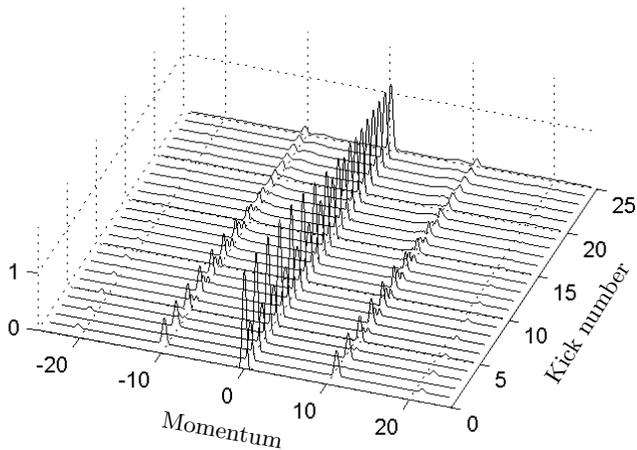}
\end{center}
\caption{Momentum distribution versus kick number for the numerical
simulation in Fig.~\protect\ref{fig:datatheoryline}}
\label{fig:momdist}
\begin{picture}(0,0)
\put(-60,66){\begin{rotate}{-10}{\small Momentum} \end{rotate}}
\put(90,82){\begin{rotate}{58}{\small Kick number} \end{rotate}}
\end{picture}
\end{figure}

The damping in the oscillation of the energy is due to dephasing associated
with the finite width of the condensate's initial momentum distribution. The
initial momentum state is not perfectly reconstructed after each free
evolution period because different momentum components of the initial
distribution have a slightly different Talbot time (associated with their
slightly different phase evolution). This is illustrated in Fig.~\ref{fig:momdist}.
 The rate of coupling between momentum states is not uniform across the
momentum distribution of the condensate. The central (zero momentum) region
of the initial condensate momentum distribution couples to the higher-order
momentum states at a slower rate than the non-zero wings of the condensate
wave function. This causes, for example, the development of the
double-peaked structure in the first order diffraction components. As time
evolves the cycling between momentum states for different components of the
initial distribution become progressively out of phase.

\begin{figure}[tbh]
\begin{center}
\includegraphics[width=\linewidth]{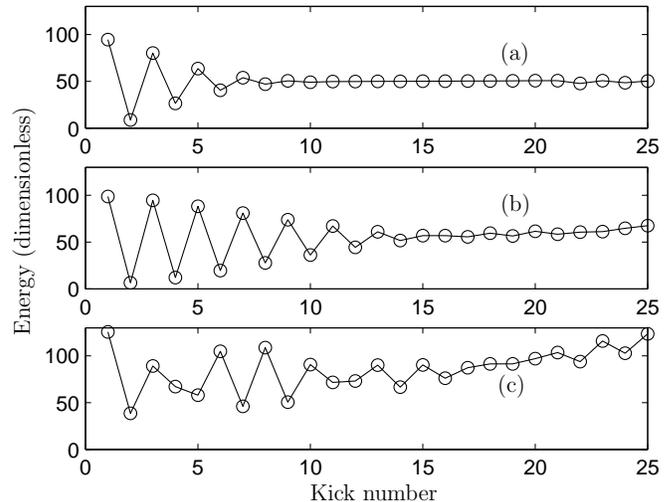}
\end{center}
\caption{Numerical simulations of energy versus kick number for the quantum
anti-resonance condition, $\kbar=2\pi$.
(a)  No collisional interaction ($C=0$), (b) collisional interaction
corresponding to the experiment ($C=50$), and (c) strong collisional
interaction ($C=1000$).}
\label{fig:theoryplot}
\end{figure}

This process of dephasing occurs even in the absence of collisions. Figure~\ref{fig:theoryplot} illustrates the results of
theoretical simulations, showing the energy dynamics with and without interactions.
For the experimental condition of $C=50$ (Fig.~\ref{fig:theoryplot}(b)), the
behavior is essentially the same as the collisionless case (Fig.~\ref{fig:theoryplot}(a)) although
dephasing occurs on a longer timescale. This is due to the fact that the
initial condensate momentum width decreases with increasing collisional
interactions \cite{Pethick2004}.

After a significant time one might expect a rephasing of the condensate
wavefunction, leading to quasi-periodic dynamics. While we have not observed
this, our numerical simulations indicate that some rephasing is possible for the
collisionless case, but this is highly sensitive to noise, and we predict
that rephasing will not occur in the nonlinear regime.

In Fig.~\ref{fig:theoryplot}(c) the nonlinear term $C$ in our calculation
is a factor of 20 times larger than that corresponding to our experiment,
and the behavior is no longer dominated by the damped oscillations. We
begin to predict what appears to be unstable behavior similar to that
predicted by Zhang \textit{et al.} \cite{Zhang2003}, and this is consistent
with our values of kick strength and nonlinearity. Experimentally this is
attainable with condensates of $>10^7$ atoms (which is beyond
our reach), or by increasing the s-wave scattering length via a Feshbach
resonance \cite{Marte2002}. Although Feshbach resonances have not been
observed in the magnetically trapped states of $^{87}$Rb, and one would
therefore need to use other spin states confined in an optical dipole trap,
this is a particularly appealing method for controlling the strength of the
nonlinearity. Another possibility for reaching the chaotic regime is by
using a much lower kick strength. We repeated our measurements for a kick
strength $\kappa =4.125$ and observed similar features to those presented in
Fig.~\ref{fig:datatheoryline}, with the main difference being a smaller amplitude of the energy
oscillations. We estimate that we would have to reduce our kick
strength by a factor of 100 to enter the chaotic regime predicted by Zhang
\textit{et al.} \cite{Zhang2003}. While it may seem straightforward to simply further reduce the
intensity of the kicking beams, this reduces the energy of the system to the
point where shot-to-shot variations exceed the predicted signal. For a kick strength lower than $\kappa\approx 4$ the signal to noise is compromised and the energy of our system becomes immeasurable.

In summary, we experimentally investigated the possibility of using
nonlinear collisional interactions in a typical Bose-Einstein condensate to
observe chaotic dynamics in the quantum-kicked harmonic oscillator system.
We applied a pulsed, far-detuned, optical standing wave to a rubidium Bose
condensate, and measured the system energy as a function of kick number for
the case of the quantum anti-resonance condition at $\kbar=2\pi$. We found that, even in the presence of nonlinear interactions, our system exhibits the well-known periodic behavior
associated with the linear system. Using numerical solutions to the
Gross-Pitaevskii equation, we showed that observed dephasing of the
oscillations is due to the finite width of the condensate's initial momentum
distribution. This severely limits the possibility of observing an extended
period of chaotic behavior in the energy of the system.

\acknowledgments The authors acknowledge the support of the Marsden Fund of
the Royal Society of New Zealand (grant 02UOO080) and KJC acknowledges the
support of FRST Top Achiever Doctoral Scholarship (TAD 884). We thank Simon
Gardiner and Scott Parkins for helpful discussions.


\end{document}